\documentclass[12pt,preprint]{aastex}

\usepackage{lscape}

\begin{document}

\title{Dissecting the region of 3EG J1837$-$0423 and HESS J1841$-$055 with INTEGRAL}

\author{V. Sguera\altaffilmark{a}, G.E. Romero\altaffilmark{b,c}, A. Bazzano\altaffilmark{d}, N. Masetti\altaffilmark{a}, 
        A. J. Bird\altaffilmark{e}, L. Bassani\altaffilmark{a}.}
\altaffiltext{a}{IASF/INAF Bologna, via Piero Gobetti 101, 40129 Bologna, Italy}
\altaffiltext{b}{Instituo Argentino de Radioastronomia, CCT La Plata-CONICET,  C.C.5, (1894) Villa Elisa, Buenos Aires, Argentina}  
\altaffiltext{c}{Facultad de Cs. Astronomicas y Geofisicas, UNLP; Paseo del Bosque S/N, (1900) La Plata, Argentina}
\altaffiltext{d}{IASF/INAF Roma, via Fosso del Cavaliere 100, 00133 Roma, Italy}
\altaffiltext{e}{School of Physics and Astronomy, University of Southampton, SO17 1BJ, UK}

\begin{abstract}
3EG J1837$-$0423 and HESS J1841$-$055 are two unidentified and peculiar high-energy sources located in the same region of the sky,
separated by $\sim$ $1.4\deg$.  Specifically, 3EG J1837$-$0423 is a transient MeV object detected by EGRET only once
during flaring activity that lasted a few days while  HESS J1841$-$055 is a highly extended TeV source.  
We attempted to match  the high-energy emission from the unidentified sources 
3EG J1837$-$0423 and HESS J1841$-$055 with X-rays (4--20 keV) and soft $\gamma$-rays (20--100 keV) 
candidate counterparts detected through  deep INTEGRAL observations of the sky region. As a result we propose the SFXT AX J1841.0$-$0536 
as a possible candidate counterpart of 3EG J1837$-$0423, based on spatial proximity and transient behavior. Alternatively, AX J1841.0$-$0536 
could be responsible for at least a fraction of the entire TeV emission from the extended source HESS J1841$-$055, based on a striking spatial correlation.
In either case, the proposed association is also supported  from an energetic standpoint by a theoretical scenario where AX J1841.0$-$0536 is a low magnetized 
pulsar which, due to accretion of massive clumps from the supergiant companion donor star, undergoes sporadic changes to transient Atoll-states where a magnetic tower 
can produce transient jets and as a consequence  high-energy emission. In either case (by association with  3EG J1837$-$0423 or alternatively with  HESS J1841$-$055), 
AX J1841.0$-$0536  might be the prototype of a new class of Galactic transient MeV/TeV emitters.

\end{abstract}
\keywords{gamma-rays: observations, X-rays: binaries, X-rays: observations}


\section{Introduction}

Soft-gamma ray astronomy is a relatively young research field which experienced a golden age in the last decades.
Breakthroughs have been achieved thanks to $\gamma$-ray satellites  carrying instruments such as CGRO/EGRET, INTEGRAL/IBIS and Swift/BAT
whose survey capabilities unveiled the extreme richness of objects in the  soft $\gamma$-ray sky. 
Recently, ground-based very high-energy (VHE) $\gamma$-ray astronomy  has also shown rapid progress 
with important results reported by the  third generation of imaging atmospheric Cherenkov telescopes such as HESS, MAGIC, VERITAS and CANGAROO.
A rapidly growing list of $\sim$ 50 firmly identified sources have been detected at TeV energies, including
AGNs ($\sim$15), pulsar wind nebulae ($\sim$18), supernova remnants ($\sim$10) and X-ray binaries ($\sim$ 4). In addition,  
there are $\sim$ 24 TeV sources still unidentified with no firm  counterpart at other wavelengths.

Among the different types of TeV sources, $\gamma$-ray binaries are rapidly becoming a subject 
of topical and major interest in  soft-gamma ray astronomy. The four firm TeV binaries detected so far are 
systems hosting a bright massive OB star as companion of the compact object, namely 
LS 5039 (Aharonian et al. 2005a), LS I+61 303 (Albert et al. 2006), PSR B1259$-$63 (Aharonian et al. 2005b) and Cygnus X-1 (Albert et al. 2007).
Their VHE emission provides evidence that particles can be efficiently accelerated to energies as high as $\sim$ 30 TeV.
Different scenarios have been proposed in the last few years to explain the emission mechanism at such high energies from HMXBs.  
Some are based on the microquasar accretion/jet framework, with both leptonic and hadronic scenarios.
In the former the $\gamma$-ray emission is due to inverse Compton scattering between relativistic electrons in the jet and seed stellar and/or synchrotron photons 
(Bosch-Ramon et al. 2006, Paredes et al. 2006, Dermer \& Boettcher 2006) whereas in the latter the radiation  originates from inelastic proton-proton interactions 
between relativistic hadrons from the jet and cold protons or nuclei from the stellar wind (Romero et al. 2003, 2005).
Proton-photon interactions are also an interesting possibility for both low and high mass microquasars (see Romero \& Vila 2008). 
An alternative leptonic scenario for the origin of $\gamma$-rays takes into account the region of interaction between the relativistic wind of a young
neutron star and the wind of the stellar companion (Maraschi \& Treves 1981, Tavani \& Arons 1997a, Dubus 2006; see Romero et al. 2007a 
for a comparison between the pulsar wind and microquasar models).
Finally, variable hadronic $\gamma$-ray emission could be produced in transient High Mass X-ray Binaries (HMXBs) through the Cheng-Ruderman mechanism in the
magnetosphere of an accreting neutron star (Orellana et al. 2007a).

In this paper, we report on new INTEGRAL imaging data of the sky region containing the two still unidentified $\gamma$-ray sources
3EG J1837$-$0423 and HESS J1841$-$055. Our goal is to find their best candidate counterparts, to this aim we discuss their 
spatial and temporal relationship with the nearby sources detected by INTEGRAL in X-rays (4--20 keV) 
and soft $\gamma$-rays (20--100 keV).


\section {The MeV/TeV emitting region}

3EG J1837$-$0423 and HESS J1841$-$055 are two unidentified and peculiar high-energy sources located in the same region of the sky. 
Their angular separation is $\sim$ $1.4\deg$ (see Fig. 1).

HESS J1841$-$055 is an unidentified TeV source discovered by HESS during the Galactic Plane survey (Aharonian et al. 2008). It shows 
a highly extended ($\sim$ 24$'$ semi-major axis) and  possibly bipolar morphology in its TeV image. 
The spectrum is best fitted by a power law with photon index $\Gamma$ $\sim$ 2.4, the  
flux is $\sim$  5.8$\times$10$^{-11}$  erg cm$^{-2}$ s$^{-1}$  (0.5$-$80 TeV) with a corresponding  luminosity of  
$\sim$ 3.3$\times$10$^{35}$  erg s$^{-1}$, assuming a distance of $\sim$ 6.9 kpc (see section 4.3.2).
The extended TeV morphology suggests that HESS J1841$-$055 is the blend of more than one source and to date no obvious unique counterpart 
has been found at other wavelengths for the entire source. 
Aharonian et al. (2008) searched for possible  counterparts using standard catalogs of sources thought to be responsible of VHE emission, 
i.e. pulsar wind nebulae (PWN), supernova remnant (SNR), HMXBs. By doing so, they found four candidates each of which could be responsible for at least 
part of the entire TeV emission:  the two pulsars PSR J1841$-$0524 and  PSR J1838$-$0549, the diffuse source G26.6$-$0.1 which is a candidate SNR based on its ASCA spectrum 
and finally the HMXB AX J1841.0$-$0536.

3EG J1837$-$0423 is an unidentified EGRET source  (E$>$100 MeV) listed in the  third EGRET catalog 3EG (Hartman et al. 1999) with coordinates and 
95\% probability contour radius equal to RA=$279.4\deg$, DEC=$-4.4\deg$ and $\sim0.5\deg$, respectively.
Recently  Casandjian \& Grenier (2008) reported a revised version of the 3EG catalog based on
the reanalysis of the whole EGRET dataset by using a new and much improved galactic interstellar 
emission model based on very recent dark gas, CO, HI, and interstellar radiation field data. 3EG J1837$-$0423 is listed in such revised version of the 
3EG catalog  with  a bigger 95\% probability contour radius ($\sim0.7\deg$)  and very similar coordinates (RA=$279.6\deg$, DEC=$-4.34\deg$)
with respect to those reported in  the 3EG catalog. 3EG J1837$-$0423  is a very peculiar transient source discovered  in 1995 
(viewing period 20--30 June) during a very bright $\gamma$-ray flare lasting only a few days (Tavani et al. 1997b) and reaching a 
peak flux above 100 MeV of $\sim$ 6.4$\times$10$^{-10}$  erg cm$^{-2}$ s$^{-1}$. The corresponding  $\gamma$-ray luminosity is  
$\sim$ 3.6$\times$10$^{36}$  erg s$^{-1}$, assuming again a distance of $\sim$ 6.9 kpc (see section 4.3.2). 
When active, 3EG J1837$-$0423 was the second brightest $\gamma$-ray source in the sky.
The photon spectrum above 30 MeV during the peak emission is best fit by a power law with spectral index equal to $\sim$ 2.1.
EGRET pointed in the direction of 3EG J1837$-$0423 fifteen more times during a period of $\sim$ 4 years and  
each observation was typically 1-2 weeks long; the source was never significantly detected again except
during the viewing  period 18--25 July 1994 and only with a marginal significance ($\sim$ 3$\sigma$). 
Such marked transient behavior  is strongly reminiscent of a blazar, but the 99\% error circle of 3EG J1837$-$0423 contains no radio object at a flux level 
consistent with other blazars seen by EGRET (Tavani et al. 1997b). 

As can be clearly seen in  Fig.1,  HESS J1841$-$055 and 3EG J1837$-$0423 are closely located in the sky. 
At first glance, their considerably large positional uncertainty regions could misleadingly suggest a spatial association  and 
a common nature. However, this is implausible  because of their completely different 
high-energy characteristics (HESS J1841$-$055 is a highly extended and non variable source while  3EG J1837$-$0423 is a point-like and 
transient source), moreover the chance probability
of positional coincidence between all galactic HESS sources and EGRET objects is as high as $\sim$ 10\% (Funk et al. 2008).
Our goal is to find their best candidate counterpart and to this aim we consider HESS J1841$-$055 and 3EG J1837$-$0423
as two distinct sources, not physically associated.
In the following sections we report a study of the field  containing  both high-energy sources, using data obtained 
with INTEGRAL in the energy bands  4--20 keV and 20--100 keV.



\section{INTEGRAL observations of the MeV/TeV emitting region} 

\subsection{Data analysis}

INTEGRAL data collected with JEM-X (Lund et al. 2003) and IBIS (Ubertini et al. 2003) have been considered in this work.
The data reduction was carried out with the release 7.0 of the Offline Scientific Analysis (OSA) software.
INTEGRAL observations are typically divided into short pointings called Science Windows (ScWs) of $\sim$ 2000 seconds duration.
Through the paper, the spectral analysis was performed  using \emph{Xspec}  version 11.3 and all spectral uncertainties are given at the 90\% confidence level
for a single parameter of interest.  

\subsection{INTEGRAL Imaging}

Because of the regular monitoring of the Galactic Plane by  INTEGRAL, the sky region including the two unidentified high-energy sources 
HESS J1841$-$055 and 3EG J1837$-$0423 is now well covered by JEM--X (4--20 keV) and IBIS (20--100 keV), with total exposure times 
of $\sim$  160 ks and  $\sim$  3 Ms respectively.

Fig. 1  shows the 20--100 keV IBIS significance mosaic map with superimposed 
the error region of HESS J1841$-$055 (ellipse) and 3EG J1837$-$0423 (50\% to 99\% probability contours as taken from the 3EG catalog and 95\% error circle
as taken from the revised 3EG catalog).
We note that AX J1841.0$-$0536  is the only soft $\gamma$-ray source
to be detected within the HESS error ellipse, the other likely candidates proposed 
by Aharonian et al. (2008) are not visible in the IBIS map and their estimated 2$\sigma$ 
upper limits are  $\sim$ 0.2 mCrab (or 1.5$\times$10$^{-12}$  erg cm$^{-2}$ s$^{-1}$) and
$\sim$ 0.4 mCrab (or 3.7$\times$10$^{-12}$  erg cm$^{-2}$ s$^{-1}$) in the 20--40 and 40--100 keV energy bands, respectively. 
As for  3EG J1837$-$0423, no soft $\gamma$-ray sources have been detected inside its uncertainty contours, however we note that 
two objects have been detected in its immediate nearness: AX J1841.3$-$0455 and AX J1841.0$-$0536.

Moreover we took into account the possible contribution,
inside or in the proximity of EGRET and HESS error regions,  from other catalogued soft $\gamma$-ray sources not detected in the IBIS mosaic in Fig. 1. 
To this aim, we used the  INTEGRAL reference catalog which  classifies previously known X-ray and $\gamma$-ray sources
that  are, or have been at least once, brighter than $\sim$1 mCrab above 3 keV and then expected to be detected by INTEGRAL.
The cross correlation with the  HESS uncertainty region resulted in only one  catalogued object, AX J1841.0$-$0536, which is visible in Fig. 1.  
Concerning the EGRET uncertainty regions,  the cross correlation resulted in three catalogued objects:
AX J1841.0$-$0536, AX J1841.3$-$0455 (both visible in Fig. 1) and GS 1839$-$04 which is not detected above 20 keV
but is nevertheless indicated in Fig.1 with a square.

Next, we searched the entire public data archive of JEM--X1 (from revolution 171 to 528) and JEM--X2 (from revolution 46 to 170) for pointings where  
HESS J1841$-$055 and 3EG J1837$-$0423 were  within the fully coded FOV of JEM$-$X ($\sim$$2.4\deg$).
As a result,  a total of  24 ScWs (JEM--X2) and 105 ScWs (JEM--X1) were selected,  spanning the viewing periods from 
11 March 2003 to 16 October 2003 and  10 March 2004 to 2 September 2006, respectively. We used all available ScWs 
to generate a mosaic significance map in the 4--20 keV band with a total exposure of $\sim$ 130 ks (JEM--X1) and $\sim$ 30 ks (JEM--X2). 
Fig. 2 shows the JEM--X significance map having the longest exposure (JEM--X1) again with the superimposition of the HESS J1841$-$055 and 
3EG J1837$-$0423 uncertainty regions. Apart from  AX J1841.0$-$0536 and  AX J1841.3$-$0455,  there is another JEM--X detection ($\sim$ 5$\sigma$ level) at 
RA=18 42 40.8 DEC=-04 23 41.9 (error radius $\sim$ 3$'$). This position is fully compatible with that of the catalogued X-ray source 
AX J1842.8$-$0423 (RA=18 42 48 DEC=-04 23 00, error radius $\sim$ 3$'$). Moreover, Fig. 2  shows that AX J1842.8$-$0423 is very likely associated 
with the INTEGRAL reference catalog source GS 1839$-$04 (error radius $\sim$ 24$'$).


\section{Dissecting the MeV/TeV emitting region}
As stated in the previous section,  three sources were detected by INTEGRAL during deep observations of the MeV/TeV emitting region:
AX J1841.0$-$0536, AX J1842.8$-$0423, AX J1841.3$-$0455. Only one of them, AX J1841.0$-$0536,  is located inside the uncertainty region of 
HESS J1841$-$055: the spatial association is striking being right at the center of the TeV error ellipse. On the contrary, no object has 
been detected by INTEGRAL inside the 95\% error circle  of 3EG J1837$-$0423 ($\sim$ $0.7\deg$ radius).  Then, we 
looked for counterparts in a bigger error area having a radius of  $\sim$ $1.2\deg$, i.e. almost twice the 95\% error circle radius.  
We are aware that this is a dangerous approach because of the possibility 
that unrelated sources could be included, however as pointed out by Thompson et al. 
(1995) and Hartman et al. (1999), the position uncertainty regions provided in the 3EG 
catalog are statistical only: additional systematic 
position errors due to inaccuracies in the Galactic diffuse radiation model and source confusion are strongly recommended to the users, particularly for those
EGRET sources located on the Galactic Plane. As example, the corresponding counterparts of three well known gamma-ray 
pulsars (3EG J0534+2200, 3EG J0633+1751 and 3EG J0834-4511) are located well outside their 99\% EGRET probability contours (Hartman et al. 1999).
Therefore it is clearly worth searching for counterparts of unidentified EGRET sources even at large distances, i.e. well outside  their error boxes. 
In the specific case of 3EG J1837$-$0423, we point out that by using a much  improved background model (Casandjian \& Grenier 2008)
its 95\% error circle radius increased from $\sim0.5 \deg$ (3EG catalog) to  $\sim0.7 \deg$ (revised 3EG catalog). 
Inside an error region  of $\sim$ $1.2\deg$ radius, we 
note that the position of the three sources AX J1841.0$-$0536, AX J1842.8$-$0423 and AX J1841.3$-$0455 
might be still fully compatible with 3EG J1837$-$0423 in view of the systematics and they could be considered as potential counterparts.
In this section, we report results from  archival and new X-ray/soft 
$\gamma$-ray observations of AX J1842.8$-$0423, AX J1841.3$-$0455 and AX J1841.0$-$0536.
Moreover, we discuss their possible physical association with the corresponding  spatially associated high-energy source.

\begin{figure}[t!]
\begin{center}
\plotone{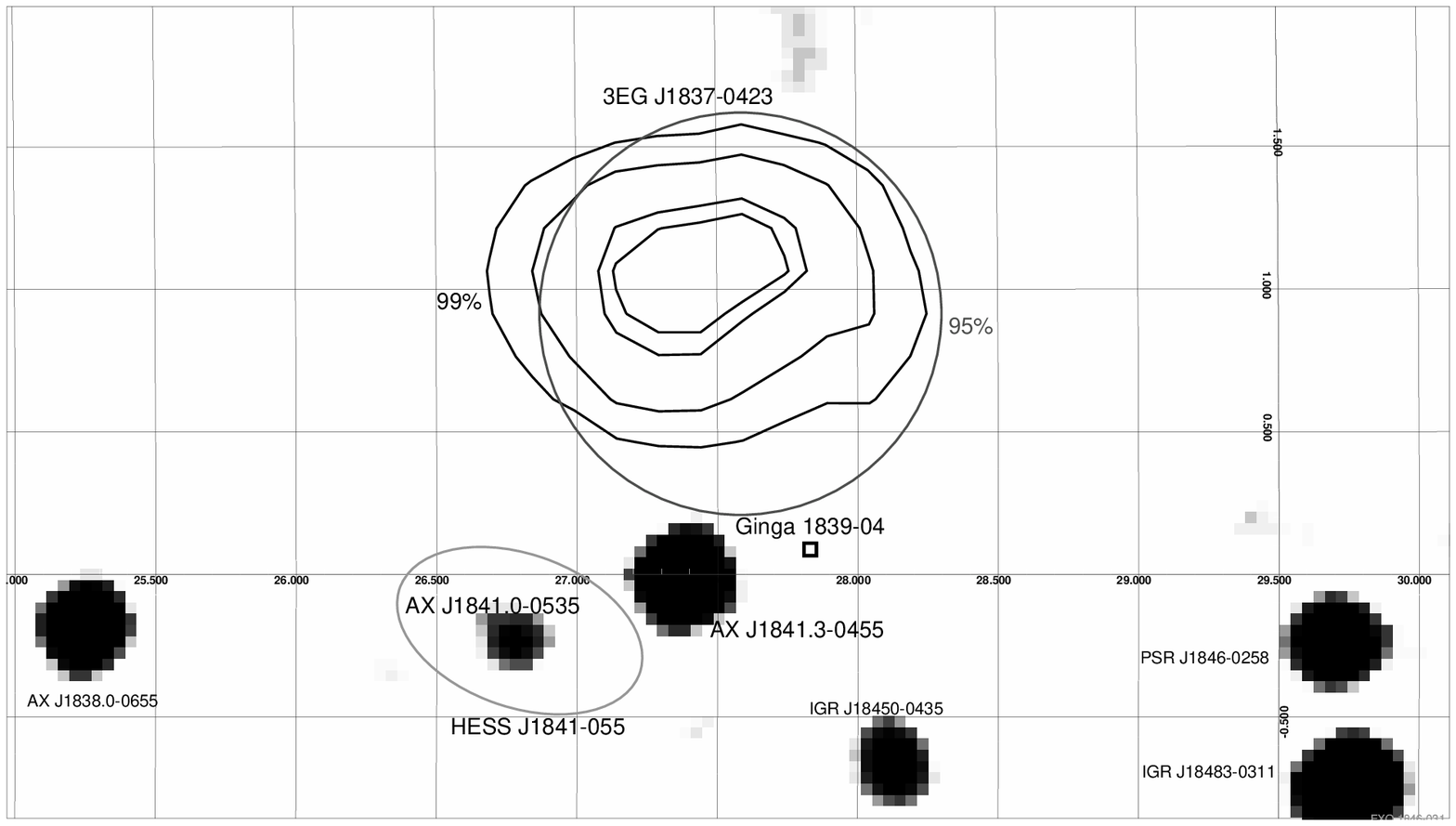}
\caption{IBIS/ISGRI mosaic significance map (20--100 keV, $\sim$ 3 Ms exposure time) of the region including the two unidentified high-energy sources 
HESS J1841$-$055 (ellipse) and 3EG J1837$-$0423 (50\%, 68\%, 95\% and 99\% probability contours as taken from the 3EG catalog and 95\% error circle as taken 
from the revised 3EG catalog). Also shown is the position of the X-ray source GS 1839$-$04 (square), not detected in the mosaic.}
\end{center}
\begin{center}
\plotone{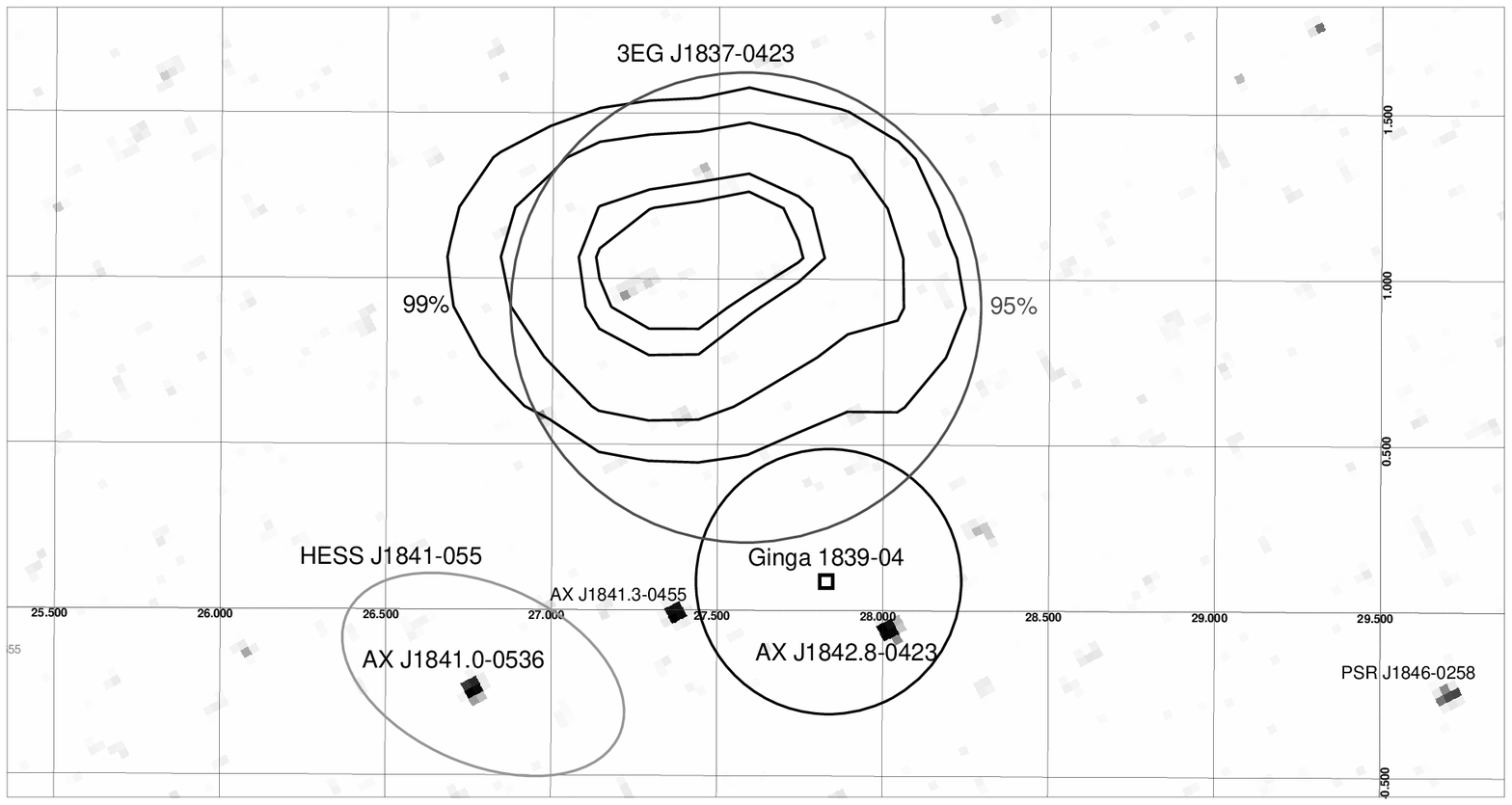}
\caption{JEM--X1 mosaic significance map (4--20 keV, $\sim$ 130 ks exposure time) of the region including HESS J1841$-$055 (ellipse) and 
3EG J1837$-$0423  (50\%, 68\%, 95\% and 99\% probability contours as taken from the 3EG catalog and 95\% error circle (bigger) as taken 
from the revised 3EG catalog). The smaller circle represents the uncertainty region  of GS 1839$-$04.}
\end{center}
\end{figure}

\subsection {AX J1842.8$-$0423 (possibly also GS 1839--04)}

AX J1842.8$-$0423 is an unidentified transient X-ray source discovered in October 1996 during a $\sim$ 35 ks observation of
ASCA while surveying the Scutum arm region (Terada et al. 1999). It was undetectable in a previous ASCA observation in 1993 and also in a subsequent one in
April 1997, providing an upper limit to the flux and outburst activity duration 
of $\sim$ 2$\times$10$^{-13}$  erg cm$^{-2}$ s$^{-1}$ (2--10 keV) and $\sim$ half a year, respectively.  
When detected,  it exhibited a 2--10 keV flux of $\sim$5$\times$10$^{-12}$  erg cm$^{-2}$ s$^{-1}$
with no variability. The ASCA spectrum is well approximated by an absorbed power law with $\Gamma$=2.9$\pm$0.4 and N$_{H}$$\sim$5$\times$10$^{22}$ cm$^{-2}$;
moreover the most interesting spectral property is the strong iron line detected at $\sim$ 6.8 keV with an equivalent width of  $\sim$ 2.4 keV.
In general, LMXBs occasionally show iron lines while HMXBs sometimes show significant Fe-K lines which do not normally
exhibit an equivalent width in excess of $\sim$ 1 keV (Nagase 1989). The overall X-ray behavior of AX J1842.8$-$0423
is indeed quite peculiar and intriguing; Terada et al. (1999, 2001) proposed that it may be explained in terms of
a close binary involving a magnetized white dwarf viewed from pole-on inclination. In this case, the unusually strong 
iron line is interpreted as arising from line-photon collimation in the accretion column 
of the white dwarf, as a result of resonance scattering of line photons. Such possibility is empirically supported by X-ray observations of some 
magnetic white dwarf binaries classified as AM Herculis-type (Ishida et al. 1998, Misaki et al. 1996).

As stated in the end of  section 3.2, the detection in the JEM--X1 significance mosaic map (Fig. 2) at  RA=18 42 40.8 DEC=-04 23 41.9 (error radius $\sim$ 3$'$)
is fully compatible with the position of AX J1842.8$-$0423.  Unfortunately, insufficient statistics did not allow us 
to extract a meaningful JEM--X spectrum and light curve. The 4--20 keV flux was estimated as $\sim$ 3$\times$10$^{-12}$ erg cm$^{-2}$ s$^{-1}$, i.e. 
similar to that measured by ASCA in 1996. We point out that  AX J1842.8$-$0423 was not detected in the JEM--X2 mosaic, supporting its transient X-ray behavior.
Moreover, a deeper inspection of the total JEM--X1 dataset (spanning the time interval from 10 March 2004 to 2 September 2006) revealed that the longest interval
of detectability of the source was $\sim$ 5 months (from 18 March 2006 to 2 September 2006), i.e. a similar duration activity as constrained  by ASCA. 
It is also worth noting that AX J1842.8$-$0423 falls in the much larger error circle ($\sim$ 24$'$ radius) of the Ginga source GS 1839$-$04 (Fig. 2), 
which is the only catalogued X-ray source present in  a $\sim$ 15$'$ radius circle region around AX J1842.8$-$0423.  
GS 1839$-$04 was discovered by Ginga about two decades ago (Koyama et al. 1989, 1990), during an outburst lasting several days; since then no 
more X-ray detections have been reported in the literature by other X-ray missions. The Ginga X-ray spectrum was fit by an absorbed 
power law with $\Gamma$=1.9$\pm$0.2 and X-ray flux of $\sim$6$\times$10$^{-12}$ erg cm$^{-2}$ s$^{-1}$ (2--10 keV). 
The optical counterpart of  GS 1839$-$04 is still unknown,  however the X-ray source is  reported in the latest catalog of HMXBs by Liu et al. (2007), 
mainly because of the  possible discovery by Ginga of $\sim$ 81 seconds pulsation, although at $\sim$ 4.5$\sigma$ level. 
We think that the  HMXB scenario is uncertain but note that the spatial correlation, the transient X-ray behavior and  the X-ray spectral shape 
and flux, all seem to suggest that  GS 1839$-$04 is very likely AX J1842.8$-$0423.

AX J1842.8$-$0423 is located far away from the uncertainty region of HESS J1841$-$055 but it is close to the 95\% probability contour of 3EG J1837$-$0423.
At first glance, the spatial proximity and the transient X-ray behavior would make  AX J1842.8$-$0423 a possible candidate counterpart of the transient 3EG J1837$-$0423.
However, we consider such an association highly unlikely on the basis of the following findings: AX J1842.8$-$0423 is very likely a magnetized white dwarf binary system; 
it has been significantly detected at soft X-rays (4--20 keV) but not above 20 keV; 
its transient X-ray activity (timescale of several weeks) is significantly longer 
than that recorded by EGRET (timescale of few days).

\subsection {AX J1841.3$-$0455}

AX J1841.3$-$0455 is an anomalous X-ray pulsar (AXP) located at the center of the the small 
($\sim$4$'$ diameter) SNR Kes 73. AXPs are rare objects closely concentrated along the Galactic Plane (see Kaspi  2007 for a review).
Their ''anomalous'' X-ray luminosities ($\sim$ 10$^{33}$--10$^{35}$ erg s$^{-1}$, 2--10 keV) are orders of magnitude too high to be explained 
by rotational energy release due to spin down. On the contrary, the so-called magnetar model, based on the decay of very strong magnetic fields (10$^{14}$--10$^{15}$ G), 
is able to explain the observed characteristics of AXPs. Traditionally, AXPs
were considered soft X-ray sources (0.5--10 keV) with thermal-like spectra (kT$\sim$0.4--0.7 keV) plus a steep power law component ($\Gamma$$\sim$3--4).
Recently, INTEGRAL has changed this traditional view by detecting bright and persistent hard tail emission 
described by power law with $\Gamma$$\sim$1--1.5 and no sign of 
a break up to $\sim$ 150 keV (Kuiper et al. 2004, 2006). However, since no counterparts have been found in the MeV domain by
COMPTEL (0.75--30 MeV) and EGRET (E$>$100 MeV), these spectra are expected to show breaks.  

Different types of X-ray flux variability are displayed by AXPs: from moderate flux changes
up to a factor of few on time scales of years  to intense burst activity lasting milliseconds to few hours (Kaspi et al. 2007). Specifically,   
AX J1841.3$-$0455 is known to be a stable AXP (Kaspi et al. 2007), no bursting activity has been recorded over $\sim$ 20 years of observations with 
Ginga, ASCA, RXTE and  BeppoSAX. We investigated the long term IBIS light curve  (20--100 keV) of the source, spanning a time interval from March 2003 to  
April 2006, and we  confirm the persistent hard X-ray emission with no sign of flaring activity. Moreover,
we performed a spectral analysis of the IBIS spectrum (20--200 keV) which is best fit by a power law 
with a hard photon index ($\Gamma$=1.55$\pm$0.1), i.e. very similar to that reported by Kuiper et al. (2004) using RXTE data. 

As we can note from the INTEGRAL significance mosaics in Fig. 1 and 2, AX J1841.3$-$0455 has been significantly detected both by IBIS
and JEM--X. It is located far away from HESS J1841$-$055 (well outside its error ellipse) so it should not be considered
as its candidate counterpart. On the contrary,  it is close to the 95\% error circle of 3EG J1837$-$0423.  
However,  we can confidently assert that AX J1841.3$-$0455 is not physically associated 
to the unidentified EGRET source in the light of their very different X-ray/soft $\gamma$-ray  behaviors as well as  of the findings on AXPs reported above.

\subsection {AX J1841.0$-$0536}

\subsubsection{Archival X-ray observations}

AX J1841.0$-$0536 (also known as  IGR J18410$-$0536) is a 4.7 seconds transient X-ray pulsar 
discovered by ASCA in 1994 and then detected again in 1999; in both cases it showed a fast X-ray flaring activity
with flux increasing from $\sim$ 10$^{-12}$ erg cm$^{-2}$ s$^{-1}$ to $\sim$ 10$^{-10}$ erg cm$^{-2}$ s$^{-1}$ (2--10 keV) 
within only $\sim$ 1 hour (Bamba et al. 2001). Subsequently, no more X-ray flares were reported in the literature until 
the launch of INTEGRAL which detected three fast hard X-ray flares having a duration of a few hours and a 20--80 keV peak flux  of  
$\sim$ 10$^{-9}$ erg cm$^{-2}$ s$^{-1}$ (Rodriguez et al. 2004, Sguera et al. 2006).  

A 20 ks Chandra pointed observation in 2004 detected the source during a phase   
of no major flaring activity (Halpern et al. 2004),  the flux level was $\sim$ 4$\times$10$^{-12}$ 
erg cm$^{-2}$ s$^{-1}$ (0.5--10 keV) likely representing the quiescent X-ray emission.
Since October 2007, Swift/XRT has been performing a monitoring campaign of AX J1841.0$-$0536  (Sidoli et al. 2008) and   the source is usually
detected with a low level X-ray activity of  $\sim$ 3.5$\times$10$^{-12}$ 
erg cm$^{-2}$ s$^{-1}$  (2--10 keV). Finally, we performed a cross correlation of the Chandra error circle of AX J1841.0$-$0536  with all 
radio catalogs available in the HEASARC database. This has resulted in no catalogued objects, suggesting that  AX J1841.0$-$0536
is not a radio emitter, however a deep pointed radio observation is needed to fully confirm such possibility.

\subsubsection{Optical properties: reddening and distance}
Thanks to the Chandra very accurate position, Halpern et al. (2004) identified the optical counterpart 
of AX J1841.0$-$0536  with  USNO-A2.0 0825\_12601262, a reddened star with weak H$_\alpha$ in emission (Halpern et al. 2004). 
Recenlty, it has been classified as a B1\,Ib-type supergiant through IR spectroscopy (Nespoli et al. 2008); 
this allowed the  classification of  AX J1841.0$-$0536 as a member of the newly discovered class of Supergiant Fast X-ray transients 
(Negueruela et al. 2006; Sguera et al. 2005, 2006).

We estimated reddening and distance for this SFXT, considering its average optical 
and NIR absolute magnitudes (Halpern et al. 2004) and colors of an early-type B1\,I star (Lang 1992; Wegner 1994). 
We found that the reddening in the optical $V$ band along the source line of sight is $A_V \sim$ 6 mag, applying the Milky Way extinction law by Cardelli et al. (1989).
This figure, using the empirical formula of Predehl \& Schmitt (1995), translates into a N$_H$ value of $\sim$1.1$\times$10$^{22}$ 
cm$^{-2}$ along the line of sight of AX J1841.0$-$0536. When compared with the N$_{\rm H}$ value derived 
from our X--ray spectral fitting (see section 4.3.3), this suggests the presence of additional absorbing material in the vicinity of the X--ray source, 
likely due to the accretion stream flowing onto the compact object in this X--ray system.
Using the absolute and the observed $R$ magnitudes for this
source (again assuming that the companion is a B1\,I supergiant star) and the estimate of the absorption in the 
optical-NIR bands, we infer a distance to this source of $\sim$6.9$\pm$1.7  kpc; this is consistent with the hypothesis that AX J1841.0$-$0536
lies in the Sagittarius arm tangent of the Galaxy,
possibly on the side closer to Earth, given the
relatively low amount of $A_V$ compared to the Galactic
one along the source line of sight ($\sim$53 mag, according to Schlegel et al. 1998). 
We point out that our retrieved value for the distance is compatible, within the uncertainties, with that reported by Nespoli et al. (2008).

\begin{figure}[t!]
\plotone{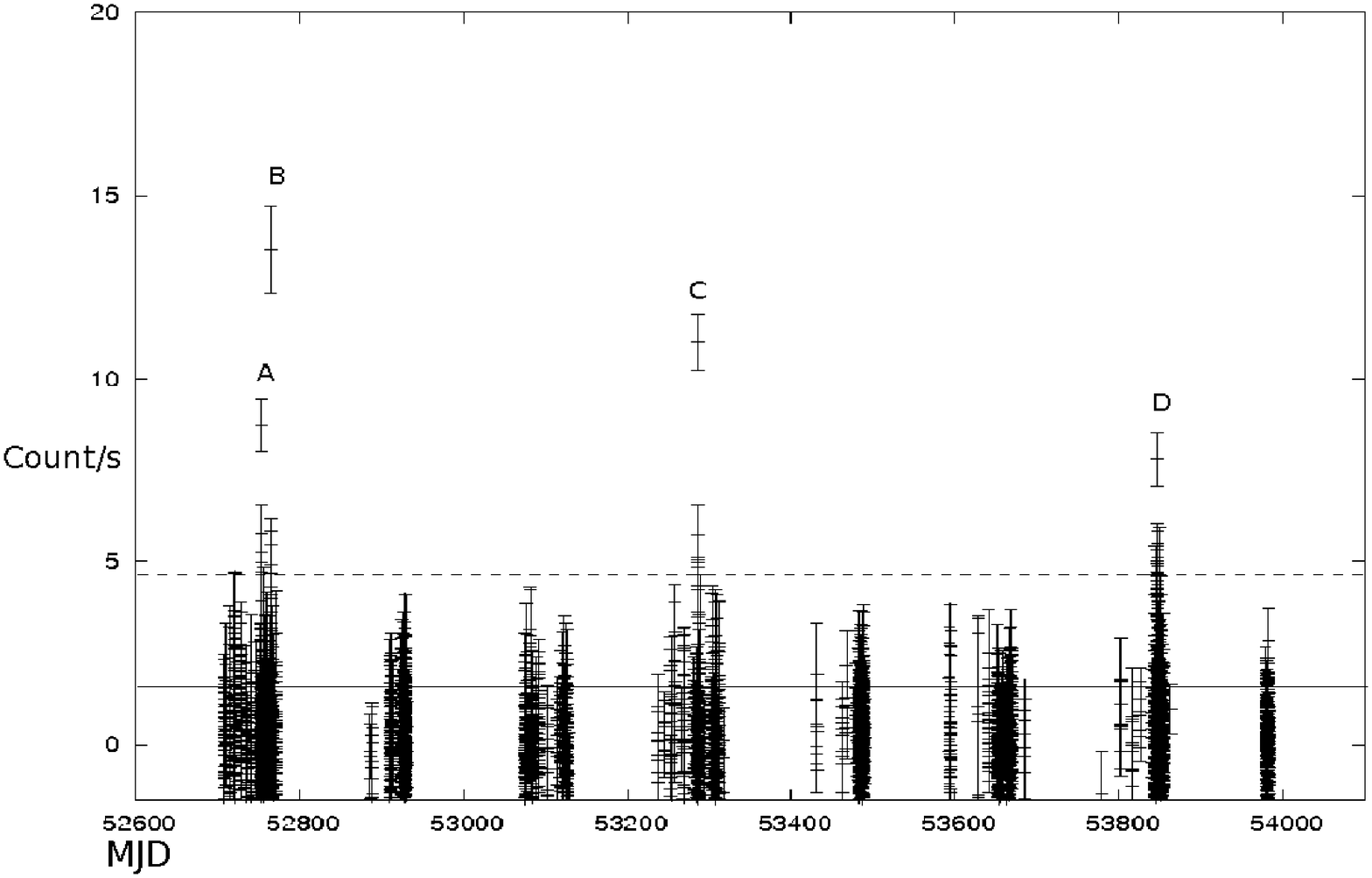}
\caption{ISGRI long term light curve (20--60 keV) of AX J1841.0$-$0536. Time and flux axis are in MJD and count s$^{-1}$, respectively.
Each data point represents the average flux during one ScW ($\sim$ 2000 seconds).}
\end{figure}
\begin{figure}
\plotone{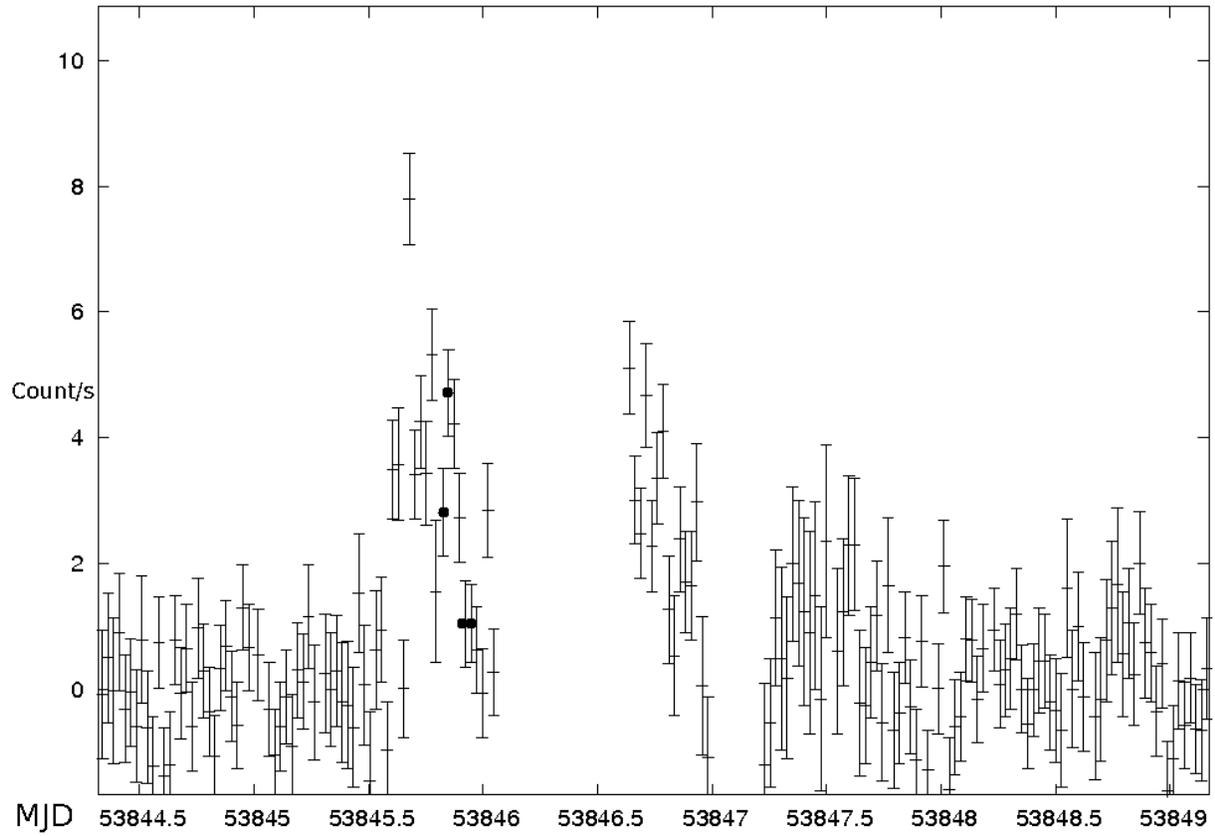}
\caption{Zoomed view of the ISGRI light curve of flare D in Fig. 3. Time and flux axis are in MJD and count s$^{-1}$, respectively.
Each data point represents the average flux during one ScW ($\sim$ 2000 seconds). The black dots indicate the ScWs during which 
the source was also in the JEM-X FOV.}
\end{figure}

\begin{figure}[t!]
\plotone{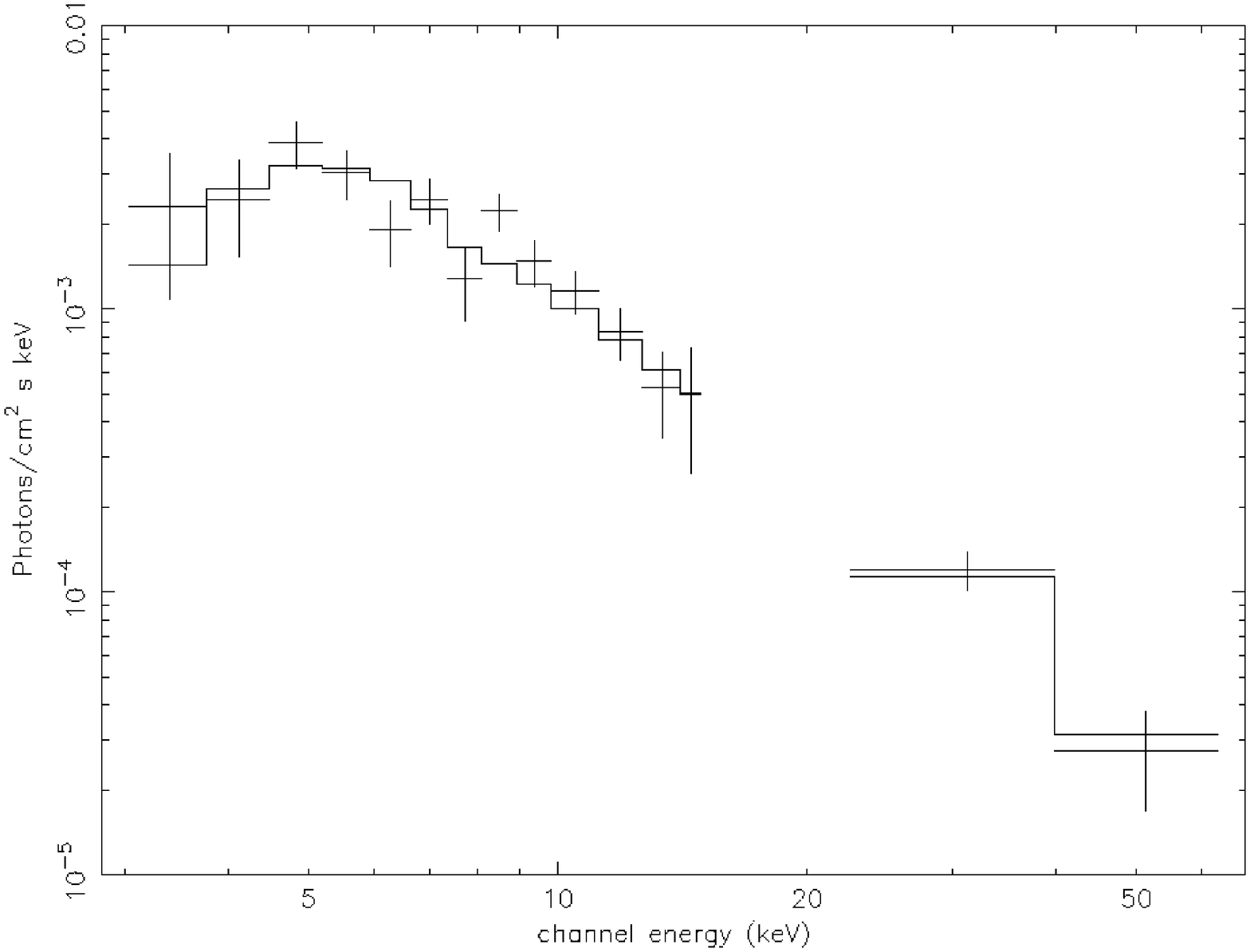}
\caption{Unfolded combined JEM--X and ISGRI spectrum (3--60 keV) of AX J1841.0$-$0536 during flare D in Fig. 3 and 4.}
\end{figure}

\subsubsection{New JEM--X and  IBIS/ISGRI results}  

We searched the entire  public IBIS dataset
(from end of February 2003 to end of September  2006)
for new outburst activity from  AX J1841.0$-$0536.  Fig. 3  shows its  long term 20--60 keV IBIS light curve on the ScW
timescale with  the flux extracted from each pointing where the source was within $12\deg$ of the center of the FOV. 
The black line  represents the 2$\sigma$ upper limit at the ScW level ($\sim$10 mCrab or 1.2$\times$ 10$^{-10}$ erg cm$^{-2}$ s$^{-1}$) and 
we note that most of the time AX J1841.0$-$0536 is in quiescence showing only rare signs of flaring activity above a flux 
of $\sim$  3$\times$10$^{-10}$ erg cm$^{-2}$ s$^{-1}$ (dashed line in Fig. 3). In fact, above this level four fast X-ray flares are evident in 
the light curve. The flares labeled as A, B and C have already been reported in the literature (Rodriguez et al. 2004, Sguera et al. 2006) with 
very similar duration (a few hours) and peak flux ($\sim$ 10$^{-9}$ erg cm$^{-2}$ s$^{-1}$, 20--80 keV). The flare D is newly discovered and 
Fig. 4 displays a zoomed view of its light curve. 
In spite of the gap in the middle due to visibility constraints,  it is evident that the 
source underwent an outburst longer than usual, with a total duration of $\sim$ 2.5 days.  
Initially  the flux was consistent with zero and suddenly it flared up to $\sim$ 50 
mCrab or 6$\times$10$^{-10}$ erg cm$^{-2}$ s$^{-1}$ (20--60 keV) in just $\sim$ 2 hours, then   
it dropped again to a very low level. Although the gap prevented  a full coverage of the outburst activity,
we can reasonably assume that another fast X-ray flare, of which we could see  the decay phase, took place during the gap. 
Unfortunately the statistical quality of the IBIS/ISGRI data is insufficient to perform a pulsations search  during the flaring activity.

The IBIS/ISGRI spectrum of the total outburst activity  (22--60 keV) is best described by a  power law with a steep photon index 
$\Gamma$=3.0$\pm$0.35 ($\chi^{2}_{\nu}$=1.45, 13 d.o.f.) or alternatively  by a 
thermal Bremsstrahlung model with kT=19$^{+5}_{-4}$ keV ($\chi^{2}_{\nu}$=1.44, 13 d.o.f.). 
Because of the different JEM--X/ISGRI FOV, we have low energy coverage of the outburst during only four ScWs
indicated in Fig. 4 by black dots;  their extracted JEM--X spectrum (3--15 keV) was best fit by an absorbed  power law
with  $\Gamma$=2.1$\pm$0.6 and  N$_{H}$=14$^{+18}_{-12}$$\times$10$^{22}$ cm$^{-2}$. We note that the latter exceeds the Galactic absorption along the line of sight
($\sim$ 1.9$\times$10$^{22}$ cm$^{-2}$). Unfortunately, the N$_{H}$ could not be well constrained because the JEM--X data extend down only to 3 keV, i.e. 
not low enough in energy to fully allow an investigation of the absorption. 
We also performed the broad band spectral analysis of the simultaneous JEM--X/ISGRI flare spectrum: the best fit ($\chi^{2}_{\nu}$=1.1, 125 d.o.f) 
was achieved by an absorbed power law (see Fig. 5) with $\Gamma$=2.5$\pm$0.6, 
N$_{H}$=23$^{+19}_{-14}$$\times$10$^{22}$ cm$^{-2}$ and a cross calibration constant of 0.8$^{+0.8}_{-0.3}$.

AX J1841.0$-$0536 was also detected at $\sim$ 5$\sigma$ level (4--20 keV)
in the JEM--X1 mosaic (Fig. 2) and at $\sim$ 4$\sigma$ level in the JEM--X2 mosaic.
With the aim of measuring its quiescent X-ray emission, we intentionally excluded from the JEM--X mosaic analysis those very few ScWs during which the source was detected 
in outburst. The 4--20 keV flux from both JEM--X1 and JEM--X2 detections is $\sim$ 2$\times$10$^{-12}$ erg cm$^{-2}$ s$^{-1}$,  
such measurement likely represents  the quiescent state of the source; indeed it is very similar
to other values previously reported in the literature (Halpern et al. 2004, Sidoli et al. 2008).
If we assume  a distance of  $\sim$6.9 kpc then the corresponding X-ray luminosity is $\sim$ 1.1$\times$10$^{34}$ erg s$^{-1}$ and the dynamic
range of the source is $\sim$ 10$^{3}$. Unfortunately, the insufficient JEM--X statistics did not allow us 
to extract a meaningful spectrum and light curve of this quiescent state.

\subsubsection{Association with 3EG J1837$-$0423 or  alternatively with  HESS J1841$-$055}

AX J1841.0$-$0536 is characterized by a  striking spatial association  with  HESS J1841$-$055, 
however its X-ray/soft gamma-ray behaviour is completely different 
than that of the HESS source (point like and transient nature versus extended and non variable nature). 
Conversely, AX J1841.0$-$0536 and 3EG J1837$-$0423 share a similar fast transient behaviour while their spatial 
match is not so precise\footnote{Positions of EGRET sources, being model background-dependent, should be considered only as indicative (Casandjian \& Grenier 2008).}
Despite these draw backs, it is still worth studying
the possible physical association of AX J1841.0$-$0536 with   HESS J1841$-$055 or alternatively with  3EG J1837$-$0423.

In the case  of HESS J1841$-$055, AX J1841.0$-$0536 is the only X-ray (4--20 keV) and soft $\gamma$-ray (20--100 keV) object detected by INTEGRAL 
within the HESS error ellipse.
It is located only $\sim$ 3$'$ from the HESS coordinates, i.e. right at the center of the TeV ellipse. Its transient X-ray behavior  as well as  point-like nature
do not agree with the extended TeV emission seen in  HESS J1841--055, however it is still reasonable to postulate that AX J1841.0$-$0536
could be responsible for at least a fraction of the entire TeV emission.
In principle, the  three other catalogued objects located inside the HESS 
uncertainty region (the two radio pulsars PSR J1841--0524 and PSR J1838--0549 and the candidate SNR G26.6$-$0.1) could  contribute to the 
remaining components of the TeV emission as it is largely known that PWN systems and SNRs are a prominent class of TeV galactic sources.
In fact, $\sim$  10 SNRs have been detected at TeV energies and they show a clear extended TeV morphology. 
With its angular size of $\sim$ 10$^{'}$, the candidate SNR  G26.6$-$0.1 could possibly contribute to a component of the VHE emission from the extended
TeV source HESS J1841--055.  Although G26.6$-$0.1 was not detected at soft gamma-rays (20--100 keV) by IBIS, this should not be taken as a strong proof 
element to exclude it as a possible counterpart because IBIS is not particularly suited for the detection of extended sources such as  G26.6$-$0.1.
As for PWN systems,  they are generally associated with  young and energetic pulsars;
to date $\sim$ 20 PWNe have been associated to point-like TeV sources and almost all of them have also been detected  at soft gamma-rays by IBIS.
In our specific case, the two pulsars have not been detected  by INTEGRAL in both energy bands 4--20 keV and 20--100 keV, 
despite very deep observations ($\sim$ 160 ks and $\sim$ 3 Ms, respectively).  Moreover,  no catalogued PWN is  associated with the pulsars
and, if taken separately, each would require an impossible efficiency of $\sim$ 200\%  (Aharonian et al. 2008) to explain
the TeV emission. All this casts doubts on the potential TeV nature of the two pulsars.
Finally,  it is very intriguing that a HMXB system with a neutron star and a supergiant companion (such as  AX J1841.0$-$0536)  
is  located  right at the core of the TeV source HESS J1841$-$055, although we took into account the 
possibility that such an association could be simply a chance coincidence. To this aim we calculated the probability of finding a supergiant HMXB, 
such as  AX J1841.0$-$0536, inside the HESS error ellipse by chance. Given the number of supergiant HMXBs detected by IBIS 
within the galactic plane (Bird et al. 2007), defined here as restricted to a latitude range of $\pm$5$^{\circ}$, we estimated a probability of $\sim$0.4\% 
i.e. 0.15 chance coincidences are expected. Such a very low probability may suggest a real physical association  between AX J1841.0$-$0536 and  HESS J1841$-$055.

In the case of  the transient 3EG J1837$-$0423, AX J1841.0$-$0536 is the only X-ray and soft $\gamma$-ray source 
detected in its immediate nearness and whose characteristics and behavior suggest a likely  association. In fact, 
AX J1841.0$-$0536 spends the majority of the time in quiescence
and very occasionally undergoes fast X-ray transient activity 
with a typical duration of a few hours, rarely a few days, i.e. similar to the transient activity of 3EG J1837$-$0423 recorded by EGRET.  
Therefore we suggest that AX J1841.0$-$0536 is the best candidate counterpart to the peculiar and elusive $\gamma$-ray transient 3EG J1837$-$0423.
As we previously reported  in the case of HESS J1841$-$055, we calculated the probability of finding by chance a 
supergiant HMXB inside the 3EG J1837$-$0423 error circle with a radius of $\sim$1$^\circ$.2.
The estimated probability is equal to $\sim$5.4\%,  such value is dominated by the considerably large chosen area (radius $\sim$1$^\circ$.2); in fact 
if we consider the smaller area pertaining to the 95\% probability contour (radius $\sim$0$^\circ$.7)  the chance probability drops to a lower value of 
$\sim$ 1.8\%. In summary,  given the above numbers, we conclude that the association of AX J1841.0$-$0536 with  3EG J1837$-$0423 is 
possibly real although a chance coincidence cannot be excluded.


\section{Origin of MeV flares or  extended TeV emission from AX J1841.0$-$0536}

Although the flaring behavior of both AX J1841.0$-$0536 and 3EG J1837$-$0423 along with the spatial proximity makes it tempting to postulate a physical relation, 
it remains open whether there is a physical mechanism to support the identification. The mechanisms proposed for 
the known TeV binaries and microquasars mentioned in the Introduction cannot be applied here, since in all those cases the sources are not 
X-ray transients. Systems like LS I +61 303, LS 5039, and PSR B1259$-$63 are not only persistent high-energy emitters, but also periodically 
variable sources, where the variability is modulated by the orbital period. Whatever produces the flares in AX J1841.0$-$0536 seems 
to have an intrinsically sporadic character.    

It has been suggested that the fast flares of SFXTs like AX~J1841.0$-$0536 are due to the interaction of the magnetized neutron star with clumps 
in the wind of the supergiant companion donor star (in $'$t Zand 2005, Leyder et al. 2007, Walter \& Zuritas Heras 2007, Negueruela et al. 2008). These clumps seem 
to be a common feature in the winds of hot stars (e.g. Owocki \& Cohen 2006). The characteristics of the clumps are not well established, but densities of 
the order of $\rho\sim 10^{-12}$ g cm$^{-3}$ and radii of $R_{\rm cl}\sim 10^{11}$ cm are likely (e.g. Romero et al. 2007b). Assuming a spherical shape, 
the mass of the clumps can be $M_{\rm cl}\sim 4 \times 10^{21}$ g. The clumps will be accreted onto the neutron star only if the magnetospheric radius, $R_{\rm M}$, 
is less than the co-rotation radius, $R_{\rm \Omega}$; otherwise centrifugal forces will expel the matter (Davidson \& Ostriker 1973, Stella et al. 1986):
\begin{equation}
R_{\rm \Omega}>R_{\rm M}. \label{R-rel}
\end{equation}

The magnetospheric radius is obtained by balancing the matter pressure to the magnetic field pressure, i.e. $\rho V^{2}=B(R_{\rm M}^{2})/8\pi$. 
Since $B(R_{\rm M})=B_{\rm NS} R_{\rm NS}^{3}/R_{\rm M}^{3}$, this yields:
\begin{eqnarray}
R_{\rm M}&=&2.6 \times 10^{6} \left(\frac{\rho_{\rm cl}}{\rm g\; cm^{-3}}\right)^{-1/5} \left(\frac{B_{\rm NS}}{10^{12}\; \rm G}\right)^{2/5}  
\nonumber \\& & \times  \left(\frac{M_{\rm NS}}{M_{\odot}}\right)^{-1/5}  \left(\frac{R_{\rm NS}}{10^{6}\; \rm cm}\right)^{6/5}\;\;\rm cm, \label{R_M}	
\end{eqnarray}
where $R_{\rm NS}$ is the radius of the neutron star, $M_{\rm NS}$ its mass and $B_{\rm NS}$ its surface magnetic field. The subscript `cl' refers 
to the clump. In estimating Eq. (\ref{R_M}) we considered that the infall velocity is $V=(2G M_{\rm NS}/r)^{1/2}$ (e.g. Massi \& Kaufman Bernad\'o 2008).

The co-rotation radius is given by:
\begin{equation}
R_{\rm \Omega}= \left(\frac{G M_{\rm NS}P^{2}}{4\pi^{2}}\right)^{1/3}.	
\end{equation}
Here, $P=2\pi/\Omega$ is the spin period. For a typical neutron star mass of $1.4 M_{\odot}$ and the observed period of 4.7 s for AX J1841.0$-$0536 (see Sect. 4.3.1) we get:
\begin{equation}
R_{\rm \Omega}=4.7 \times 10^{8} \;\; \rm cm. 
\end{equation}

Fast X-ray flares with a peak luminosity of $L_{\rm X}\sim 10^{37}$ erg s$^{-1}$ can be produced if the clump material impacts onto the surface 
of the neutron star. If a flare has a duration of the order to $\Delta t \sim 10^{4}$ s, the accretion rate from the clump matter will be 
$\dot{M}\sim M_{\rm cl}/ \Delta t \approx 4\times 10^{17}$ g s$^{-1}$. Around 10 per cent of the rest mass energy of the accreted material is released as luminosity:
\begin{equation}
	L_{\rm X}\approx 0.1 \dot{M} c^{2}\;\;\rm erg \; s^{-1}\sim 3.6 \times 10^{37}\;\;\rm erg \; s^{-1}.
\end{equation}
This means that the sporadic interaction with massive clumps can explain the observed X-ray flares {\sl if } the matter can reach the surface 
of the neutron star. However, relation (\ref{R_M}) imposes important constraints if we take into account the required energy budget. Since the infall 
velocity is determined by the mass of the  neutron star which cannot depart too much from the canonical value of $1.4 M_{\odot}$, we are left 
with the sole possibility that the surface magnetic field of the pulsar should be $B_{\rm NS}\leq 1.8 \times 10^{12}$ G. Systems with long periods 
of $\sim 1000$ s can accommodate even magnetars (Stella et al. 1986, Bozzo et al. 2008).  

In the quiescent state, the X-ray luminosity of such system  seems to be unusually high, $\sim 10^{34}$ erg s$^{-1}$. The density contrast between the clumps and 
the background wind can reach values of $10^{3-5}$ (Runacres \& Owocki 2005). If we adopt the lower value, i.e. $\rho_{\rm cl}/\rho_{\rm wind}\approx 10^3$, 
then the quiescent accretion rate will be $\dot{M}\sim M_{\rm wind}/ \Delta t \approx 4\times 10^{14}$ g s$^{-1}$. The quiescent X-ray luminosity results, therefore, 
$L_{\rm X}\sim 10^{34}\;\;\rm erg \; s^{-1}$ in accordance with the observations (see Sect. 4.3.3). The requirement of 
$R_{\rm \Omega}>R_{\rm M}$ then imposes $B_{\rm NS}\leq 1.6\times 10^{10}$ G, according to Eq. (\ref{R_M}) and the value of $R_{\rm \Omega}$. Then, 
in order to explain both the flaring and quiescent states as accretion onto a magnetized neutron star from a structured wind with clumps embedded in 
a background flow of low density, we need a low magnetic field in the star surface: $B_{\rm NS}\leq 10^{10}$ G.

Is there any room in this scenario for MeV-GeV $\gamma$-ray flares or extended TeV emission? Walter (2007) suggests that protons could be accelerated by 
multiple scattering of Alfv\'en waves in or close to the accretion column and then interact with material at the magnetospheric radius producing $\gamma$-rays 
through inelastic $pp$ collisions and the subsequent decays. He estimates maximum Lorentz factors for the protons of $\sim 10^8$, i.e. energies of $\sim 10^{17}$ eV. 
This suggestion is based on the work by Smith et al. (1992). In these calculations only synchrotron proton losses are taken into account. However, the particle density 
in the acceleration region during the flares (the only occasion when the energetics is sufficient to sustain $\gamma$-ray luminosities of $10^{36}$  erg s$^{-1}$) is huge, 
of the order of $\sim 10^{21}$ cm$^{-3}$. In addition, the physical conditions in the polar column are rather extreme, with ``temperatures'' of $\sim 20$ keV, as 
indicated by the ISGRI observations (Sect. 4.3.3), a photospheric emission area of $\sim 1$ km$^{2}$, and photon densities of $\sim 10^{24}$ cm$^{-3}$ (Arons 1987). 
Under such conditions $pp$ and $p\gamma$ losses are catastrophic. The proton cooling time scale through inelastic collision with thermal protons is: 
\begin{equation}
	[t^{(pp)}_{p}]^{-1}=\frac{1}{E_{p}}\frac{dE_{p}}{dt}=n_{p}c \sigma_{pp} K^{(pp)},\label{t_pp}
\end{equation}
where $E_{p}$ is the proton energy, $n_{p}$ the density of thermal protons, $K^{(pp)}\approx 0.5$ the inelasticity (fraction of proton energy lost per interaction) 
and $\sigma_{pp}\sim 35$ mb is the $pp$ cross section at GeV energies. Then, $t^{(pp)}_{p}\approx 2 \times 10^{-6}$ s. 

Synchrotron losses are given by:
\begin{equation}
[t^{(\rm synchr)}_{p}]^{-1}=\frac{4}{3}\left(\frac{m_{e}}{m_{p}}\right)^{3} \frac{c\sigma_{\rm T} B^{2}}{8\pi m_{e} c^{2}}	\gamma_{p}.
\end{equation}
where  $\sigma_{\rm T}$ is the Thomson cross section and the other symbols have their usual meaning. Then, for $B=10^{10}$ G, we have 
$t^{(\rm synchr)}_{p}=0.6 \gamma^{-1}_{p}$ s. This means that for protons of 1 TeV $pp$ losses are $3\times 10^{2}$ times more important than synchrotron losses. 
The acceleration surely takes place at some distance from the surface of the pulsar, thus this is an absolute upper limit.    

The acceleration rate of particles in the accretion column is (e.g. Begelman et al. 1990):
\begin{equation}
	[t^{(\rm acc)}_{p}]^{-1}=\frac{\eta c e B}{E_{p}},	
\end{equation}
where $\eta$ is the acceleration efficiency, which depends on the shock velocity and the mean free path of the particles. Typical velocities for turbulent 
motions in the accreting column are $\sim 10^{7}$ cm s$^{-1}$ (Smith et al. 1992). Then, under the most favorable assumptions (diffusion in the Bohm limit), 
the efficiency is $\eta\sim10^{-7}$. The maximum energy allowed for protons, taking into account the $pp$ losses, results in $E^{\rm max}_{p}\sim 1$ GeV, in such 
a way that the protons are barely relativistic. 

A most promising approach to generate relativistic particles during the accretion of the clump is the formation of a transient magnetic-tower jet that could 
carry away a fraction of the accreting material (Kato 2007). Magnetohydrodynamic simulations of the magnetic interaction between the neutron star field and 
the accreting material show that jets driven by magnetic pressure are formed along the rotation axis of the disk (Kato et al. 2004). This occurs when the 
accreting matter reaches distances of $\sim40$ gravitational radii. In our case this means $R_{\rm M}=8.4\times10^{6}$ cm, which imposes an even more tight 
constraint onto the magnetic field: $B_{\rm NS}\sim 2.1\times 10^{7}$ G. This value is well in accordance with the values expected in Atoll sources known to 
produce jets like Scorpius X-1 (Massi \& Kaufman Bernad\'o 2008). However, Atoll sources are low-mass X-ray binaries. Is it possible 
for a neutron star in a HMXB to have a magnetic field of $\sim10^{7}$ G? We will briefly discuss the issue in what follows.

The accretion of matter onto neutron stars can produce strong temperature gradients that favor thermomagnetic processes and the decrease of 
the crustal conductivity, all this resulting into an accelerated magnetic field decay (e.g. Geppert \& Urpin 1994, Urpin \& Geppert 1995). 
The magnetic field in the crust of the neutron star is given by:
\begin{equation}
\frac{\partial \stackrel{\rightarrow}{B}}{\partial t}=-\frac{c^{2}}{4\pi}\nabla\times\left(\frac{1}{\sigma}\nabla\times\stackrel{\rightarrow}{B}\right)+\nabla\times(\stackrel{\rightarrow}{u}\times\stackrel{\rightarrow}{B}).\label{B}
\end{equation}

Here, 
$$\frac{1}{\sigma}=\frac{1}{\sigma_{\rm ph}}+ \frac{1}{\sigma_{\rm imp}} $$
is the total conductivity in the crystallized crustal region, and $\stackrel{\rightarrow}{u}\propto\nabla T$ is the thermomagnetic velocity that characterizes the drift of the magnetic field under the influence of the temperature gradient. The conductivity is determined by the electrons, whose main scattering mechanisms are scattering on phonons and impurities (hence the terms $\sigma_{\rm ph}$ and $\sigma_{\rm imp}$, respectively). Since we are interested in the decay of a dipolar field, we can adopt a vector potential $\stackrel{\rightarrow}{A}=(0,\;,0\;, A_{\varphi})$, with $A_{\varphi}=s(r, t) \sin \theta /r$ (we are using spherical coordinates and standard notation). Since the drift is radially directed and the thickness of the crust $\sim 0.1 R_{NS}$, we can use the plane-parallel approximation (Geppert \& Urpin 1994) with the drift velocity in the $z$-direction (i.e. outwards). Then, Eq. (\ref{B}) can be written as:
\begin{equation}
\frac{\partial s}{\partial t}=\frac{c^{2}}{4\pi \sigma}\frac{\partial^{2}s}{\partial z^{2}}-u\frac{\partial s}{\partial z},\label{s}
\end{equation}
with the condition $\partial s/ \partial z=0$ at $z=0$. Given a prescription for the temperature, Eq. (\ref {s}) can be solved for different values of $\sigma_{\rm imp}$, since $\sigma_{\rm ph}$ is determined by the temperature and the crustal density ($\sim 10^{12}$ g cm$^{-3}$). The temperature profiles are determined by the accretion rate (Fujimoto et al. 1984). The accretion rate of dense clumps dominates the total accretion in a system like AX J1841.0$-$0536, as we have shown at the beginning of this section. Even with a very low duty cycle of $\sim 1$ \%, we have an average accretion rate of $\dot{M}\sim 10^{-10}$ $M_{\odot}$ yr$^{-1}$.

For a very pure crust, where the conductivity is basically determined by phonons,  and the above mentioned accretion rate, the thermomagnetic drift is directed outwards and the crustal field expelled. The neutron star magnetic field then decays 4 orders of magnitude in $\sim 3\times10^{6}$ yr. At higher accretion rates or lower impurity content, the decay can be even larger in the same time. We conclude, then, that young neutron stars in HMXBs can, under certain conditions, have magnetic fields as low as those invoked in this paper (few times $10^{7}$ G).  We notice, however, that only the conductivity due to electron-phonon scattering depends on the temperature (which is sensitive to the accretion rate). The phonon conductivity decreases when $T$ increases, allowing changes in the magnetic field according to Eq. (\ref{s}). The conductivity due to electron-impurity scattering depends only on the number density of impurities in the crust. The value of $\sigma_{\rm imp}$ is given by (e.g. Urpin \& Geppret 1995):
\begin{equation}
\sigma_{\rm imp}\sim 4.2 \times 10^{21} x \frac{Z}{Q}\;\;{\rm s}^{-1}.
\end{equation}
In this expression $x$ is  relativistic parameter of the electrons ($x=p_{\rm F}/mc$, with $p_{\rm F}$ the Fermi pressure), $Z$ is the charge number of the dominant ions in the crust, and $Q$ is a parameter that characterizes the number density and charge of the impurities:
\begin{equation}
Q=\frac{1}{n_{i}}\sum_{n'} n' (Z-Z')^{2},
\end{equation}
where $n_{i}$ is the number density of the dominant background ion species of charge number $Z$, and the primed parameters refer to the interloper species of impurities, over which the summation is carried out. Depending on the purity of the  crust, the conductivity $\sigma_{\rm imp}$ can dominate hindering the decrease of the magnetic field. This seems to be the case in accreting binaries with strong fields, like AO 0535+26 and similar systems.

The magnetic loops connecting the neutron star and the disk are twisted because of the differential 
rotation. Twist injection of matter from the disk results in the expansion of the loops, which creates a magnetic tower inside which the accelerated disk material 
is collimated as bipolar jets with subrelativistic speeds ($0.1-0.2c$). Magnetic reconnection at the base of the tower can inject plasmoids and the collision of 
plasmoids of different velocities will result in shocks in the outflow. Diffusive acceleration at these shocks can accelerate particles, both protons and electrons, 
up to relativistic energies. Electrons will cool almost instantly through synchrotron radiation producing X-rays (Romero \& Vila 2008a). Actually, the conditions of 
the transient jet will not be very different from those discussed by these authors. Proton-photon interactions can produce $\gamma$-rays, secondary pairs, neutrinos 
and neutrons according to the following reactions:
\begin{equation}
p+\gamma\rightarrow p+a\pi^0+b\left(\pi^++\pi^-\right)
	\label{photomeson1}
\end{equation}  
and
\begin{equation}
p+\gamma\rightarrow n+\pi^++a\pi^0+b\left(\pi^++\pi^-\right),
	\label{photomeson2}
\end{equation}  
where $a$ and $b$ are the pion multiplicities. The decay chains for the mesons are:
\begin{equation}
\pi^+\rightarrow\mu^++\nu_\mu, \quad \mu^+\rightarrow e^++\nu_{\rm{e}}+\overline{\nu}_\mu,
	\label{piondecay1}	
\end{equation}

\begin{equation}
\pi^-\rightarrow\mu^-+\overline{\nu}_\mu, \quad \mu^-\rightarrow e^-+\overline{\nu}_{\rm{e}}+\nu_\mu,
	\label{piondecay2}	
\end{equation}
 
\begin{equation}
\pi^0\rightarrow2\gamma.
	\label{piondecay3}	
\end{equation}  

Proton-proton inelastic collisions would also yield $\gamma$-rays and secundary particles if the density of the ejected plasma is high enough:
\begin{equation}
p+p\rightarrow p+ \Delta^{+} + a\pi^0+b\left(\pi^++\pi^-\right),
	\label{pp}
\end{equation}
where $ \Delta^{+}$ is a resonance that decays yielding a leading pion that takes around 17 \% of the proton energy. 

The particle distribution for protons and those for all other types of particles (primary electrons, secondary pairs, muons and pions) can be obtained solving the corresponding transport equations:
\begin{equation}
	\frac{\partial N_{i}(E, z, t)}{\partial t}+ \frac{\partial}{\partial E}\left[\left.\frac{dE}{dt}\right|_{\rm{loss}}N_{i}(E,z, t)\right]+\frac{N_{i}(E,z, t)}{t_{\rm{esc}}}+ \frac{N_{i}(E,z, t)}{t_{\rm{decay}}}=Q_{i}(E,t).
	\label{transpeq1}
\end{equation} 
where $t_{\rm{esc}}$ is the particle escape time from the acceleration region of thickness $\Delta z$ ($t_{\rm{esc}}\approx \Delta z/v_{\rm{outflow}}$), $t_{\rm{decay}}$ is the decay time for the different particles (infinity for $e$ and $p$), $dE/dt |_{loss}$ is the sum of all losses for the type of particles considered, and $Q_{i}$ is the injection function that can be normalized in accordance to the energy budget of relativistic particles of type $i$ through:

\begin{equation}
 L_{i}=\int_{V}\mathrm{d}^3r\int_{E_{i}^{\rm{min}}}^{E_{i}^{\rm{max}}\left(z\right)}\mathrm{d}E_{i}\,E_{i}\,Q_{i}(E_{i}).
\label{norminj}
\end{equation}    
  
The injection, resulting from first order Fermi acceleration mediated by the shocks formed in the magnetic tower, will have the form of a power law $Q_p(E_i)\propto E_i^{-\alpha}$, where $\alpha$ typically lies between 1.5 and 2.2 (Malkov \& Drury 2001), depending of the shock geometry, non-linear effects, etc. The particle distributions obtained solving Eq. (\ref{transpeq1}) will be affected by the losses. Reynoso \& Romero (2009) have solved Eq. (\ref{transpeq1}) in the steady state for protons, electrons, muons and pions in conditions similar to those discussed in this paper. Romero \& Vila (2009) present the corresponding energy distributions due to all relevant processes: synchrotron radiation, IC scattering, relativistic Bremsstrahlung, $p\gamma$ and $pp$ interactions, and they include the effect of adiabatic losses as well. To solve the equations out of the steady state goes beyond the scope of the present paper.  Nonetheless, some energetic considerations are in order to show that the proposed scenario is consistent.

We can assume, in accordance to the most recent studies of the accretion/ejection coupling (K\"ording et al. 2006), that around 10\% of the accretion power is injected in the collimated outflow, and around 10\% of this power is 
converted into power of relativistic particles, as it seems to be the case in microquasars and supernova remnants (e.g. Ginzburg \& Syrovatskii 1964, Bosch-Ramon et al. 2006). This means that we would have around $4\times 10^{36}$ erg s$^{-1}$ in relativistic particles injected during 
the accretion of the clump by the neutron star. So, a normal $\gamma$-ray flare would have around several times $10^{35}$ erg s$^{-1}$ at energies $E_{\gamma}>100$ MeV. These $\gamma$-rays would be the result of $p\gamma$ interactions close to the neutron star and $pp$ interactions in the dense medium around the system 
(the column density is $\sim 10^{23}$ cm$^{-2}$ and the typical size of a massive binary system $\sim 10^{12}$ cm). Protons could be confined by 
the local magnetic field. Secondary electrons and positrons from hadronic interactions can produce additional radiation through inverse Compton and 
synchrotron mechanisms (Orellana et al. 2007b, Bosch-Ramon et al. 2008). Very intense $\gamma$-ray flares such as those detected by EGRET can result from the 
accretion of a particularly large and dense clump. During the massive accretion period, when the source goes 
through the transient Atoll state, X-ray pulsations are suppressed since the matter is directed outwards. Pulsations are detectable 
during the quiescent states when the source accretes from the interclump medium.

Both $p\gamma$ and $pp$ produce neutrons, that cannot be confined by magnetic fields and could escape from the system. Depending on their energy these 
neutrons will decay at different distances, injecting both protons and electrons in the interstellar, dense medium around the binary system:
\begin{equation}
	n\rightarrow p+e^{-}+\overline{\nu_{e}}.
\end{equation}
The most energetic neutrons can reach distances of $d=\gamma^{\rm max}_{p} (886.7 \,\rm{s})\; c \sim 1$ pc. The protons from the decay of the neutrons will 
start to diffuse into the surrounding of the source, forming an extended, non-variable, $\gamma$-ray source through the ``illumination'' in $\gamma$-rays of the 
ambient matter (e.g. Bosch-Ramon et al. 2005). The total energy deposit in the medium by these neutrons during a flare will be $\sim 10^{36}$ erg s$^{-1}$,
which will be transferred to the decay products. If the duty cycle of the source is a few percent, then, on average, around $\sim 10^{34}$ erg s$^{-1}$ 
will be injected. If particles have a spectral index of $\alpha\sim 2$, the most energetic particles will interact releasing around of 10\% of their energy 
in $\gamma$-rays up to pc-scale distances, sustaining an extended high-energy source.  Notice that the protons injected by the decay 
of neutrons have a long lifetime (see Eq. \ref{t_pp}, and consider an average molecular medium of $\sim 1-10$ cm$^{-3}$). The protons  can be trapped
 by magnetic fields inside the binary system or the neighboring molecular clouds, being accumulated through many ejection episodes. Hence, 
an extended, stable source at high energies can co-exist with a transient X- and $\gamma$-ray compact source.

The above scenario is outlined only to show that the association of AX J1841.0$-$0536 with 3EG J1837$-$0423 or at least part of the emission from HESS J1841$-$055 
is possible. A more detailed and general model, where SFXT sources are considered as low-magnetized pulsars undergoing sporadic changes to an Atoll-state, due 
to the accretion of clumps from the supergiant companion, will be presented elsewhere.


\section {Discussions and conclusions}

We attempted to match, from an energetic and positional standpoint, the high-energy emission from 
3EG J1837$-$0423 and HESS J1841$-$055  with X-ray (4--20 keV) and soft $\gamma$-ray (20--100 keV) candidate counterparts 
detected through  deep INTEGRAL  observations.

In the case of HESS J1841$-$055, its TeV emission appears to have an extended morphology
which suggest contributions from several sources; the possibility that the HESS source  is the blend of more than one object has been explored.
A search, using standard catalogs of sources which could be responsible of VHE emission, led to 
four catalogued objects positionally correlated with the HESS uncertainty region: two pulsars, a candidate SNR and a HMXB.
The HMXB system AX J1841.0$-$0536 is the only object 
detected by INTEGRAL at X-ray (4--20 keV) and at soft $\gamma$-rays  (20--100 keV).
Intriguingly,  this HMXB is located at the core  of the TeV emission. 
We estimated the probability of finding the  supergiant HMXB AX J1841.0$-$0536 inside the HESS error ellipse by chance equal to $\sim$0.4\%, 
i.e. low enough not to preclude a physical association.  We are aware that the point-like nature 
and fast transient behavior of AX J1841.0$-$0536 exclude its association 
with the entire extended HESS source. Nevertheless, our study suggests that at least a fraction 
of the entire TeV emission might well be associated with  AX J1841.0$-$0536. To this aim, we presented a theoretical model supporting such a scenario.
It still remains to be understood  what sources are then contributing to the rest of the TeV emission.
It is largely known that SNRs are a prominent class of TeV galactic sources: as a consequence, the extended source and 
candidate SNR G26.6--0.1  (angular size $\sim$ 10$^{'}$) could be responsible for a  component of the entire TeV emission from   HESS J1841$-$055. 
PWN systems are also well known TeV galactic sources,  so in principle the two catalogued pulsars could be high-energy emitters
which contribute to the entire TeV emission.
However, there are the  following doubts on their potential TeV nature: $\imath$)  no catalogued PWNe are associated with the two pulsars;  
$\imath$$\imath$) when taken separately, each pulsar would require an impossible high efficiency ($\sim$ 200\%) 
to explain the VHE emission, not consistent with the range of efficiencies (0.01\%-11\%) found for other TeV PWNe; 
$\imath$$\imath$$\imath$)  none of the  two pulsars has been detected by 
INTEGRAL in the energies range  4--20 keV and 20--100 keV, despite very long observations, having in mind that 
almost all TeV PWN systems have been detected by INTEGRAL at soft gamma-rays. 
It is clear that the above informations are not sufficient to draw any definitive conclusion on which 
other sources are contributing to the entire TeV emission from  HESS J1841$-$055, apart from the HMXB AX J1841.0$-$0536.
Further and deeper multiwavelength studies are strongly needed  in order to $\imath$) support or reject the TeV nature
of the candidate SNR and the two pulsars $\imath$$\imath$) unveil the presence of still undetected 
high-energy objects (i.e. PWNe, SNRs) inside the HESS error region.

As for the peculiar transient 3EG J1837$-$0423, to date no radio or X-ray counterpart has been found inside its 95\% confidence error circle
(radius $\sim 0.7 \deg$) despite extensive searches in the past years. Bearing this in mind, we opened  the search for X-ray and soft $\gamma$-ray 
counterparts to a bigger  area, i.e. radius $\sim 1.2 \deg$.
Among the few sources detected by INTEGRAL (4--20 keV and 20--100 keV)
in the proximity of 3EG J1837$-$0423, again we identified  the transient HMXB AX J1841.0$-$0536 as the best candidate counterpart 
based on immediate nearness and similar flaring behavior. Furthermore, such association was supported from an energetic standpoint 
by proposing a theoretical mechanism able to explain the flaring MeV emission from  AX J1841.0$-$0536.
In the outlined scenario,  AX J1841.0$-$0536 is a low magnetized pulsar ($B_{\rm NS}\sim 2.1\times 10^{7}$ G)
which,  due to accretion of a massive clump
from the supergiant companion,  undergoes sporadic changes to a transient Atoll-state where a magnetic tower 
can produce transient jets. After the collision with the massive clump, everything comes back to the normal state.

One way or another (association with HESS J1841$-$055 or alternatively with 3EG J1837$-$0423), the SFXT AXJ1841.0$-$0536 
could be the prototype of a new class of Galactic transient MeV/TeV emitters. Additional evidence for the existence of such a
new class is also provided by very recent AGILE and GLAST discoveries on the Galactic Plane of several unidentified transient MeV sources lasting only a few days
(Tavani et al. 2008, Cheung et al. 2008, Pittori et al. 2008, Longo et al. 2008,  Chen et al. 2007).

Further multiwavelength observations of the entire region in radio, X-rays (i.e. XMM, Chandra and Swift/XRT), soft gamma-rays (i.e. INTEGRAL), MeV and GeV 
(i.e. AGILE and GLAST) are strongly needed in order to disentangle the emission possibilities, confirm or reject our proposed scenario  and
find a definitive counterpart to the  enigmatic and intriguing sources  HESS J1841$-$055 and  3EG J1837$-$0423.

\begin{acknowledgements}
We thank the anonymous referee for useful comments which helped us to improve the quality of this paper.
The italian authors acknowledge the ASI financial support via grant ASI-INAF I/023/05/0, I/088/06/0/ and I/008/07/0/. 
G.E.R. thanks V. Bosch-Ramon for insightful comments and the MPIfK, Heidelberg, 
for hospitality during his contribution to this work. He is supported by grants from CONICET and ANPCyT, as well as 
by the Ministerio de Educaci\'on y Ciencia (Spain) under grant AYA 2007-68034-C03-01, FEDER funds. This research has made use of 
data obtained from the HEASARC database.
\end{acknowledgements}

\end{document}